\documentclass[article,10pt,twocolumn]{IEEEtran}
\usepackage[T1]{fontenc}
\usepackage[utf8]{inputenc}
\usepackage{lmodern} 

    \usepackage{tablefootnote}
    \usepackage{adjustbox} 
    \usepackage{newtxtext,newtxmath}

    \usepackage{graphicx}
    \usepackage{booktabs}
    \usepackage{cite}
    \usepackage{adjustbox}
    \ifCLASSOPTIONcompsoc
      \usepackage[caption=false,font=footnotesize,labelfont=sf,textfont=sf]{subfig}
    \else
      \usepackage[caption=false,font=footnotesize,labelfont=small]{subfig}
    \fi
    \ifCLASSOPTIONcaptionsoff
      \usepackage[nomarkers]{endfloat}
     \let\MYoriglatexcaption\caption
     \renewcommand{\caption}[2][\relax]{\MYoriglatexcaption[#2]{#2}}
    \fi
     
    \usepackage[utf8]{inputenc}
    \usepackage{ragged2e}
    \usepackage{units}
    \usepackage{graphicx}
    \usepackage{tabulary}
    \usepackage{booktabs}
    \usepackage{mathtools}

    \usepackage{amsmath} 
    \usepackage{amsthm} 
    
    \usepackage{amssymb}
    \usepackage{amsfonts}
    \usepackage{cite}
    \usepackage{flushend}

    \usepackage{xfrac}
    \usepackage{pifont} 
    \usepackage{tabularx}
    \usepackage{comment}    
    \usepackage{setspace}
    \usepackage{xspace}
    \usepackage{acronym}
    \usepackage{multirow}
    \usepackage{upgreek} 
    \usepackage{nicefrac} 
    \usepackage{ifthen}
    \usepackage{csvsimple}
    \usepackage{hyperref}
    \usepackage{flexisym} 
    \usepackage{fp} 
    \usepackage{caption}
    \usepackage{titlecaps}
    \usepackage{optidef}
    \usepackage[thinc]{esdiff} 
    \usepackage{cleveref}
    
    \theoremstyle{definition}

    \usepackage{xargs} 
    \usepackage{xcolor}
    
    \acrodef{IoT}{Internet of Things}
    \acrodef{RV}{random variable}
    \acrodef{RAO}{random access opportunity}
    \acrodef{CIoT}{cellular IoT}
    \acrodef{LTE}{Long Term Evolution}
    \acrodef{LTE-A}{LTE-Advanced}
    \acrodef{LTE-M}{LTE-MTC}
    \acrodef{NB-IoT}{Narrowband \ac{IoT}}
    \acrodef{NR}{new radio}
    \acrodef{mMTC}{massive machine-type communications}
    \acrodef{MIMO}{multiple-input multiple output}
    \acrodef{mMIMO}{massive \acl{MIMO}}
    \acrodef{NOMA}{non-orthogonal multiple access}
    \acrodef{MAC}{medium access control}
    \acrodef{MAD}{maximum average distance}
    \acrodef{ZC}{Zadoff–Chu}
    \acrodef{FDD}{frequency division duplexing}
    \acrodef{BS}{base station}
    \acrodef{5G-NR}{5G new radio}
    \acrodef{PRACH}{Physical Random Access Channel}
    \acrodef{RACH}{Random Access Channel}    \acrodef{RA}{random access}

    \newcommand{\gformat}[1]{\mathcal{#1}}
    \newcommand{\HSymbol}{H}
    \newcommand{\V}{\gformat{X}}
    \newcommand{\E}{\gformat{E}}

    \newcommand{\HGraph}[1]{\HSymbol(\V,\E)}
    \newcommand{\Hypergraph}[2]{\HSymbol_{\code}(\V,\E)} 
    
    \newcommand{\code}{\mathcal{C}\xspace}
    
    \DeclareMathSizes{10}{9}{7}{5}
    
    \newboolean{long}   

    \usepackage{letltxmacro}
    \LetLtxMacro{\oldsqrt}{\sqrt}
    \renewcommand{\sqrt}[2][\mkern8mu]{\mkern-4mu\mathop{\oldsqrt[#1]{#2}}}

    \def\false{false}
    
    \def\finalcolor{true} 

    \ifx\finalcolor\false
        \def\revcolor{blue}
    \else
        \def\revcolor{black}
    \fi

    \newcommand\revtext[1]{\textcolor{\revcolor}{#1}}
\begin{document}
%
\author{
\IEEEauthorblockN{Giancarlo~Maldonado Cardenas, 
Diana~C.~Gonzalez, 
Judy~C.~Guevara, 
Carlos~A.~Astudillo, 
and Nelson~L.~S.~da~Fonseca
\vspace{-5mm}
\thanks{
This work was supported by the São Paulo Research Foundation (FAPESP) under grant number **/***- and **/****-*,. 
}%
\thanks{G.~Maldonado, J.~Guevara, C~A.~Astudillo, and N.~L.~S~da~Fonseca are with the Institute of Computing, University of Campinas 13083-852, Campinas, Brazil. 
D. Gonzalez is with the Pontifical Catholic University of Campinas, 13087-571 Campinas, Brazil.
(e-mails: g262946@dac.unicamp.br, diana.cristina@puc-campinas.edu.br, cguevara@unicamp.br, castudillo@unicamp.br, nfonseca@ic.unicamp.br).}%
} }

\title{\huge ML-Based Preamble Collision Detection in the Random Access Procedure of Cellular IoT Networks}%

\maketitle
\newcommand{\MYheader}{\smash{\scriptsize
    \hfil\parbox[t][\height][t]{\textwidth}{\centering {\normalsize
    Submitted to and \textit{IEEE Journal}, 2025
    }}\hfil\hbox{}}}
    \makeatletter
    \def\ps@IEEEtitlepagestyle{%
    \def\@oddhead{\MYheader}%
    \def\@evenhead{\MYheader}%
    \def\@oddfoot{}%
    \def\@evenfoot{}}
    \makeatother
    \pagestyle{headings}
    \addtolength{\footskip}{0\baselineskip}
    \addtolength{\textheight}{-0.1\baselineskip}
\begin{abstract}
Preamble collision in the Random Access Channel (RACH) is a major bottleneck in massive Machine-Type Communication (mMTC) scenarios, typical of Cellular IoT (CIoT) deployments. This work proposes a machine learning-based mechanism for early collision detection during the Random Access (RA) procedure. A labeled dataset was generated using the RA procedure messages exchanged between the users and the base station under realistic channel conditions, simulated in MATLAB.
We evaluate nine classic classifiers—including tree ensembles, support vector machines, and neural networks—across 
four communication scenarios, varying both channel characteristics (\textit{e.g.}, Doppler spread, multipath) and the cell coverage radius, to emulate realistic propagation, mobility, and spatial conditions.

The neural network outperformed all other models, achieving over 98\% Balanced Accuracy in the in-distribution evaluation (train and test drawn from the same dataset) and sustaining 95\% under out-of-distribution evaluation (train/test from different datasets).
To enable deployment on typical base station hardware, we apply post-training quantization. Full Integer Quantization reduced inference time from 2500 ms to as low as 0.3 ms with negligible accuracy loss.
The proposed solution combines high detection accuracy with low-latency inference, making it suitable for scalable, real-time CIoT applications found in real networks.

\end{abstract}
\acresetall
\vspace{-2mm}
%
\IEEEpeerreviewmaketitle
\bstctlcite{IEEEexample:BSTcontrol}
\section{Introduction}
\label{sec:Introduction}

    \IEEEPARstart{T}he rapid expansion of the use of Internet of Things (IoT) devices has driven considerable advances in various areas, such as transport, health and industrial automation \cite{iot, survey, salud}. Massive Machine-Type Communications (mMTC) stands out for enabling the simultaneous connection of a large number of devices characterised by low complexity, low power consumption and sporadic communication. This class of service is designed to support the accelerated growth of the IoT in environments with a high density of devices, and is a key component in the development of smartnets \cite{smartnetss, smartness2}. In this context, the massive IoT paradigm assumes a strategic role by enabling the distributed orchestration of devices on a large scale.
    An example of this scenario is smart cities, where low-power sensors are integrated into urban elements such as traffic lights, lampposts, vehicles, buildings and mobile devices. These sensors collect and transmit data in real time, monitoring variables such as air quality, vehicle flow, energy consumption and mobility patterns. The result is an interconnected network that provides support for automated decision-making systems.

    To address these challenges, key cellular IoT communication (CIoT) technologies such as New Radio (NR), NB-IoT, and LTE-MTC utilize the Random Access Channel (RACH)\footnote{In this paper, use RACH as a generic term for the random access channel of all CIoT technologies} for initial synchronization between user equipment (UE) and base stations (BS). The RACH process allows the BS to handle up to 64 preambles per cell, with each device randomly selecting a preamble to compete for resources \cite{3gpp}. However, this process often encounters high collision rates because multiple devices may select the same preamble simultaneously \cite{3gpp}. In scenarios with dense traffic, this issue exacerbates, increasing the likelihood of collisions and necessitating repeated random access (RA) attempts, which intensify resource contention and network congestion \cite{massive, Modeling}.



    Implementing effective preamble detection and collision resolution techniques is crucial to optimizing the throughput of the RACH procedure and conserving the scarce radio resources. Existing methods face limitations when dealing with dynamic and high-density network conditions, particularly in 5G scenarios that require rapid and reliable connectivity.

    Machine learning offers a robust alternative to overcome these limitations by providing models capable of capturing complex and non-linear patterns inherent to collision scenarios. Machine learning models, such as neural networks, can be trained to identify collisions directly from signal data, bypassing the need for explicit correlation calculations and improving the ability to adapt to diverse network conditions. Furthermore, the trained models can generalize to different channel configurations, enabling more accurate and efficient collision detection in real-time, even under high mobility or noise conditions.

    In this paper, we propose a novel machine learning-based framework designed to enhance collision detection in the RACH procedure. Our approach leverages a neural network model, selected for its superior performance in handling complex and non-linear relationships inherent in collision scenarios. The model is trained on a comprehensive dataset generated using MATLAB’s LTE System Toolbox, which simulates realistic channel conditions across three distinct scenarios. Furthermore, we apply advanced quantization techniques to optimize the neural network model, significantly reducing inference time without compromising accuracy.

The main contributions of this work are summarized as follows:
\begin{itemize}
    \item We generate labeled datasets using MATLAB’s LTE System Toolbox, simulating three channel scenarios with different Doppler spreads and coverage radii to reflect diverse mobility profiles.
    \item We evaluate and compare nine ML classifiers and select a neural network as the optimal model based on its superior Balanced Accuracy and generalization across channel conditions.
    \item We apply Dynamic Range and Full Integer Quantization to optimize model inference time, achieving a reduction of over 99\% with negligible loss in accuracy.
\end{itemize}

The remainder of this paper is organized as follows:
Section \ref{sec:Related Work} reviews existing approaches for preamble collision detection and resolution.
Section \ref{sec:proce_aleatorio} describes the system model and the signal processing steps involved.
Section \ref{sec:metodo} details the dataset generation, feature extraction, and ML model training.
Section \ref{sec:Performance} presents the experimental evaluation and performance results.
Finally, Section \ref{sec:Conclusion} summarizes the conclusions and outlines directions for future research.
    \section{Related Work}
    \label{sec:Related Work}
    Several strategies have been proposed to mitigate preamble collisions and improve the performance of the RA procedure in CIoT networks. These solutions can be broadly grouped into three categories: signal-based detection techniques, resource allocation schemes, and machine learning (ML)-based approaches.
    
    

    \subsection{Non-Standard Compliant Approaches}
    

    Similarly, the scheme proposed in~\cite{M2M} embeds the UE identifier in the preamble, allowing the base station to detect collisions before resource assignment. However, many of these methods require modifications to the physical layer or increased signaling overhead, which may limit their feasibility in standardized LTE/5G deployments.

    In this category, we include all the schemes that uses non-standard compliant methods for collision detection, such as the use of tagged preambles~\cite{early,Resolution}, sending additional signals on the PRACH. 
    For instance, the e-PACD scheme in~\cite{early} enhances correlation processing by using multiple root sequences to detective collisions during the initial stages of the RA procedure. Labeled preambles

    \subsection{Distance-based Approaches}

    Other works exploit differences in time-of-arrival (ToA) to separate overlapping transmissions via successive interference cancellation (SIC), as proposed in~\cite{non} and~\cite{Throughput}. While effective in controlled settings, these techniques often assume ideal channel conditions and introduce high computational complexity, particularly in the presence of multipath or signal overlap~\cite{delay}. 
    
    Additional approaches leverage timing information, such as time advance (TA) values~\cite{novel}, but tend to be constrained to fixed-location scenarios, limiting their applicability in mobile environments.

    \revtext{To overcome this limitation for mobile environments, the Time-Alignment-value based Random Access (TARA) approach was proposed in \cite{tara1, tara2}. In this scheme, when a detectable collision occurs, devices use the multiple TAs measured by the base station to perform an immediate new access attempt, achieving success by finding a unique TA match between the two attempts.}
    
    On another front, focusing on the challenge of massive and asynchronous access, an iterative user activity detection algorithm with delay calibration (UAD-DC) was proposed in \cite{user_activity}. Instead of a machine learning approach, this method uses an Expectation-Maximization (EM) and Compressive Sensing (Turbo-CS) framework to precisely estimate the continuous time delays of users, allowing for collision resolution from the accurate identification of multiple signals. These signal-based approaches, while effective, contrast with our proposal, which uses machine learning to extract patterns directly from the power delay profile (PDP) without the need for multiple retry steps or complex iterative models. 

    Resource allocation schemes aim to prevent or mitigate collisions by improving how preambles and uplink resources are managed. In~\cite{FNN}, a feedforward neural network is used to classify preamble states and guide non-orthogonal resource scheduling (NORS). \revtext{More recent strategies propose more holistic resource management frameworks. For instance, \cite{preamble_parallelization, qoS_aware} propose a framework that jointly optimizes the Access Class Barring (ACB) factor and preamble allocation to meet heterogeneous Quality of Service (QoS) requirements. Their work also introduces techniques such as "colliding preamble reuse," which leverages the spatial separation of devices to increase resource availability, and "preamble parallelization," where a single device transmits multiple preambles to enhance its success probability. From a different perspective, the work in \cite{grant_free_access_for_mMTC} presents an analytical approach to determine the optimal number of preambles required to keep the failure rate below a specific threshold, thereby minimizing the need for retransmissions that could overload the network. Unlike these schemes that manage access before a collision occurs or at a system-wide level, our approach focuses on detecting the collision after the preamble has been transmitted, operating directly at the physical layer.}

\subsection{ML-based Approaches}
    More recently, ML-based techniques have emerged as promising alternatives for enhancing collision detection and RA efficiency without relying on detailed channel models. Frameworks using deep neural networks (DNNs)\cite{Framework} or convolutional neural networks (CNNs)\cite{NaNA} have demonstrated strong performance in detecting preamble collisions and estimating user multiplicity. Other studies employ logistic regression and hybrid models for inferring collision counts~\cite{via}. \revtext{Following this trend, recent research has explored enriching the input information for such models. A notable example is the work by Li et al. \cite{two_dimensional}, which proposes using a Fully-Connected Neural Network (FNN) to estimate collision multiplicity in Cell-Free Massive MIMO systems. The key innovation is the use of two-dimensional information: in addition to the received signal energy (analogous to the PDP values used in our work), the model also leverages the Angle of Arrival (AOA) as a second input feature to improve detection accuracy. This contrasts with the approach presented in this paper, which focuses on extracting maximum information from a single dimension (the 24 PDP power values) but stands out for its emphasis on practical feasibility by optimizing the neural network model for ultra-low latency inference through quantization techniques, a critical aspect for implementation on real-world base station hardware.} Nevertheless, practical deployment remains challenging due to the reliance on accurately labeled datasets, often unavailable in live networks, and sensitivity to channel variability.

    Despite significant progress, most existing methods face trade-offs between detection accuracy, implementation complexity, and robustness to dynamic, high-mobility scenarios. Table~\ref{tab:table0} summarizes representative approaches across all categories, highlighting their main goals, techniques, use of machine learning, and key limitations.

\begin{table*}[!ht]
\centering
\caption{Summary of Proposed Preamble Collision Mitigation Schemes}
\label{tab:table0}
\begin{adjustbox}{width=0.95\textwidth}
\begin{tabular}{lcp{4.2cm}p{5cm}cp{6cm}}
\toprule
 \textbf{Scheme} & \textbf{Reference} & \textbf{Objective} & \textbf{Technique} & \textbf{ML} & \textbf{Limitations} \\ \midrule
 e-PACD scheme & \cite{early} & Early detection of collisions during RA & Correlation recomputation with multiple roots & No & Performance degrades in dense networks due to complex correlation patterns \\ \midrule

 PDCCH-aware RA optimization & \cite{lte_sim} & Improve RA efficiency under control channel constraints & RA scheduling and PDCCH overload control & No & Bottlenecks remain under heavy traffic \\ \midrule

 TA-based RA scheme (static devices) & \cite{novel} & Avoid collisions in fixed-location UEs & TA comparison with fixed-location knowledge & No & Not suitable for mobile devices \\ \midrule

 PACR scheme & \cite{Resolution} & Resolve preamble collisions via TA & Labeled preambles and TA discrimination & No & Limited to stationary nodes; adds complexity \\ \midrule

 NORA scheme (ToA + SIC) & \cite{non} & Separate overlapping transmissions & ToA estimation and SIC & No & Sensitive to channel quality; hard to deploy in practice \\ \midrule

 NORA + ToA-based grouping & \cite{Throughput} & Improve access efficiency in mMTC & ToA estimation, SIC, and power control & No & Requires accurate delays; computationally intensive \\ \midrule

 Delay estimation in NORA (ML) & \cite{delay} & Estimate signal delays for SIC & RTD estimation using ML with variational inference & Yes (ML inference) & High complexity; vulnerable to overlapping signals \\ \midrule

 \revtext{TARA scheme} & \cite{tara1, tara2} & Resolve preamble collisions in mobile NB-IoT scenarios & TA-value matching across two consecutive RA attempts & No & Requires a two-step retry mechanism; adds latency of one RAO. \\ \midrule

  \revtext{UAD-DC} & \cite{user_activity} & Detect active users in asynchronous massive RA & Iterative EM and Compressive Sensing (Turbo-CS) to estimate continuous time delays & No & High computational complexity from iterative signal processing. \\ \midrule

 NORS + FNN-based detection & \cite{FNN} & Classify preambles as idle, valid, or collided & FNN-based preamble classification for NORS scheduling & Yes (FNN) & Assumes accurate input data; may not generalize well \\ \midrule

 UE ID in preamble & \cite{M2M} & Detect collisions by embedding UE identity & Embedding UE ID in PRACH preamble & No & Requires physical layer changes; not backward compatible \\ \midrule

 \revtext{Access Control (QA-JACRA) Joint QoS-Aware}  & \cite{preamble_parallelization, qoS_aware} & Improve access efficiency for massive IoT with heterogeneous QoS & Joint optimization of ACB factor and preamble allocation; uses colliding preamble reuse and preamble parallelization & No & Assumes spatial separation is sufficient for preamble reuse; parallelization increases resource consumption per device. \\ \midrule

 \revtext{Analytical Grant-Free Provisioning} & \cite{grant_free_access_for_mMTC} & Minimize retransmissions by optimizing the number of available preambles & Analytical modeling of success probabilities to provision preamble resources & No & A system planning method, not a real-time resolution technique; relies on accurate traffic models. \\ \midrule

 DNN-based preamble classification & \cite{Framework} & Estimate number of UEs per preamble & DNN-based estimation of users and TA classes & Yes (DNN) & Requires labeled training data; performance varies with channel \\ \midrule

 Collision multiplicity estimation (ML) & \cite{via} & Estimate how many devices selected the same preamble & Logistic regression and neural network for multiplicity estimation & Yes (LR + NN) & Assumes availability of precise user information \\ \midrule

 CNN-based early detection & \cite{NaNA} & Detect collisions from raw IQ features & CNN-based classification of IQ signals for early collision detection & Yes (CNN) & Sensitive to noise and mobility; requires high-fidelity input \\ \midrule

 \revtext{2D-Info FNN Multiplicity Detection} & \cite{two_dimensional} & Estimate preamble collision multiplicity in Cell-Free mMIMO & FNN using both signal energy and Angle of Arrival (AOA) as input features & Yes (FNN) & Requires hardware capable of accurate AOA estimation; performance may degrade if UEs lack spatial separation. \\

\bottomrule
\end{tabular}
\end{adjustbox}
\end{table*}



Unlike previous approaches, the proposed framework provides a scalable and infrastructure-compatible solution for collision detection in massive access scenarios. While methods such as~\cite{M2M} require physical layer modifications to embed the UE ID in the preamble, our approach operates entirely at the base station side and requires no changes to the existing LTE protocol stack. Similarly, techniques like~\cite{novel} and~\cite{Resolution}, which rely on fixed time advance values or labeled preambles, are limited to stationary devices and are less effective under high mobility. In contrast, our model is trained on datasets that span diverse channel models and user mobility conditions, improving generalization to realistic CIoT environments. Furthermore, unlike correlation-based methods such as~\cite{early}, which become impractical in high-density networks due to complex correlation patterns, our approach remains robust regardless of user density. These advantages position our solution as a practical and efficient alternative for scalable collision detection in 5G and massive IoT systems.

\section{System Model}
\label{sec:proce_aleatorio}



    \label{sec:preamble-generation}
    \subsection{Zadoff-Chu Preamble Sequence} 

    In CIoT technologies, the PRACH preamble structure varies across different standards. 
    PRACH preambles are generated using Zadoff-Chu (ZC) sequences due to their favorable mathematical properties, such as constant amplitude, zero autocorrelation for cyclic shifts, and low cross-correlation between different root sequences. These characteristics ensure high orthogonality and robustness against interference and multipath effects, which are essential in dense or high-mobility environments.

    The ZC sequence of odd-length $N_{\text{ZC}}$ can be written as
    
    \begin{align}
        z_u(m) = \displaystyle\exp \left[-j\pi u\frac{m(m+1)}{\displaystyle N_{\text{ZC}}}\right],
        \label{eq:z_u}
        \end{align}
    where $u \in \{1,\ldots, N_{\text{ZC}}-1\}$ is the root index and  $m = 0,\ldots,N_{\text{ZC}}-1$ is the sample index. To generate multiple orthogonal preambles from a single root sequence, cyclic shifts are applied as follows:
    \begin{align}
        x_{u,v}(m) = z_u((m+C_v) \ \text{mod} \ N_{\text{ZC}}),
        \label{eq:x_u,v}
    \end{align}
    where $C_v=v N_{\text{CS}}$, with $v=1,\ldots, [ N_{\text{ZC}}/N_{\text{CS}}]-1$, and $N_{\text{CS}}$ is the fixed length of the cyclic shift. This value is chosen to preserve the Zero Correlation Zone (ZCZ), ensuring orthogonality among PRACH sequences even in the presence of timing uncertainty and multipath delay spread.
   
   \subsection{PRACH Waveform Generation and Reception}

   This \lcnamecref{sec:preamble-generation} describes the structure of the PRACH preamble, the transmitter and receiver architectures, and the preamble detection procedure in SC-FDMA-Based CIoT technologies.
   
   The PRACH preamble is a complex waveform composed of three segments: a cyclic prefix (CP), the Zadoff-Chu (ZC) main sequence, and a guard time (GT). The CP and GT are designed to mitigate multipath propagation effects and ensure robust time-domain synchronization at the receiver. In LTE Frequency Division Duplex (FDD) operation, the 3GPP defines four preamble formats ~\cite{UMTS}, each with different durations and CP lengths, selected based on cell size, user mobility, and deployment conditions. Formats 0 to 3 utilize 839 active subcarriers with 1.25 kHz spacing, resulting in a bandwidth of approximately 1.08 MHz and extended duration, suitable for large cells. Alternatively, Format 4, designed for small cells and low-delay spread environments, employs 139 subcarriers with 7.5 kHz spacing, occupying a narrower bandwidth of about 180 kHz and producing a shorter preamble duration.

    \begin{figure}[!b] \includegraphics[scale=0.20]{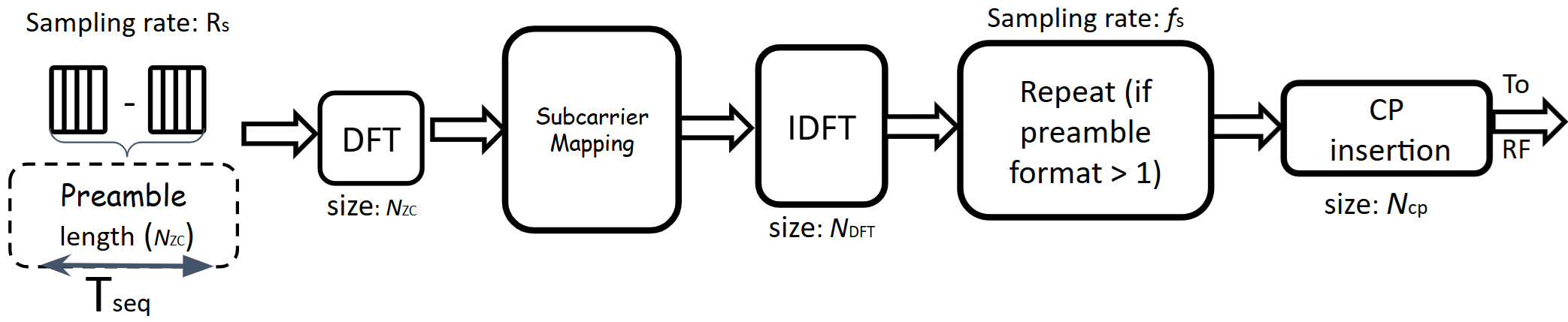}
    \centering
    \caption{Structure of the PRACH preamble transmitter}
    \centering
    \label{fig:transmitter}
    \end{figure}

    The PRACH transmission process, illustrated in Figure~\ref{fig:transmitter}, begins by transforming the time-domain ZC sequence into the frequency domain using a Discrete Fourier Transform (DFT). The frequency-domain samples are then mapped onto the designated PRACH subcarriers. Subsequently, an Inverse DFT (IDFT) is applied to convert the signal back to the time domain, after which a CP is appended to mitigate inter-symbol interference caused by multipath.

\begin{figure}[t]
    \includegraphics[scale=0.36]{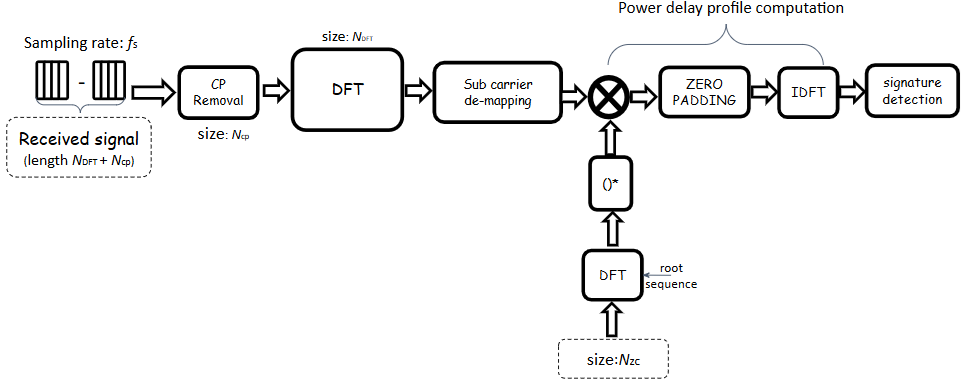}
    \centering
    \caption{Strcuture of the PRACH Receiver}
    \centering
    \label{fig:detection}
\end{figure}

    
    At the receiver, as shown in Figure~\ref{fig:detection}, the base station removes the CP and applies a DFT to the received signal. The subcarrier de-mapping block extracts the RACH preamble. Simultaneously, the BS generates a local ZC sequence, applies DFT and complex conjugation, and correlates it with the received signal in the frequency domain. The correlation vector is zero-padded and transformed back to the time domain via IDFT, yielding the Power Delay Profile (PDP).
    
\begin{figure}[!b]
    \includegraphics[scale=0.36]{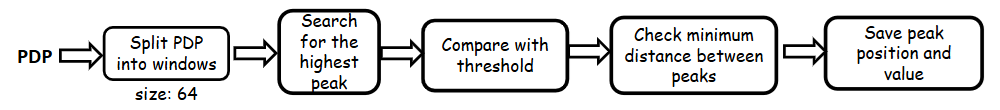}
    \centering
    \caption{Block diagram of the PDP peak detection process}
    \centering
    \label{fig:detection_peak}
\end{figure}

    Preamble detection is performed by estimating the background noise and setting an adaptive threshold using an energy-based detector. The PDP is then segmented into fixed-size windows, and peaks that exceed this dynamic threshold are identified as potential signal arrivals from active UEs. To ensure reliable detection, a minimum distance between peaks is enforced, preventing multiple detections of the same arrival. The positions and values of valid peaks are then stored, as illustrated in Figure~\ref{fig:detection_peak}.
    


    The complete PDP consists of 1,536 samples, segmented into 64 bins of 24 samples each. This segmentation enables the BS to monitor individual access attempts and identify potential collisions, typically indicated by multiple peaks within the same bin. For each detected preamble, the BS aggregates the power contributions across bins to construct a consolidated representation of the multipath environment and assess the likelihood of collisions. This fine-grained temporal resolution is essential for improving collision detection and optimizing radio resource allocation, particularly in dense IoT deployments.

\subsection{Random Access Procedure}
\label{sec:rap}
    \begin{figure*}[!t]
        \includegraphics[scale=0.25]{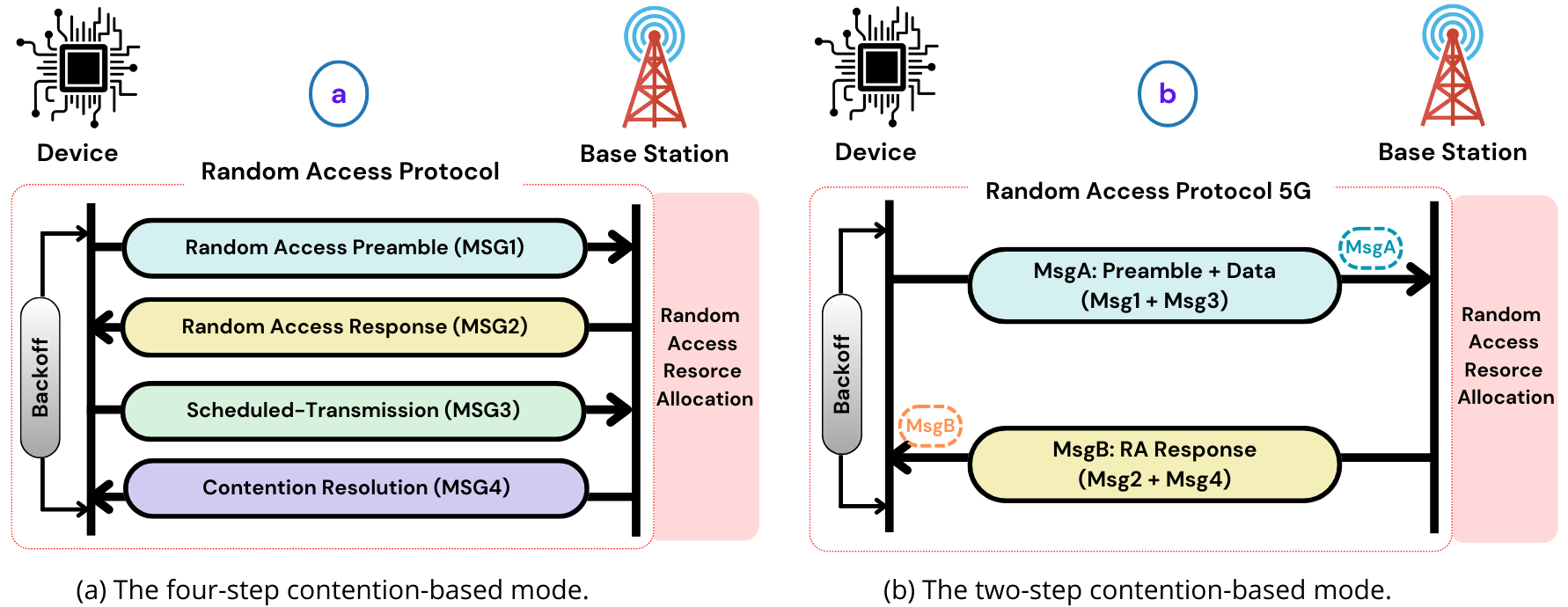}
        \centering
        \caption{the 4-step and 2-step contention-based Random Access Procedures}
        \centering
        \label{fig:steps_en}
    \end{figure*}
    
    The Third Generation Partnership Project (3GPP) specifies two operation modes for the random access procedure: a non-contention-based access, typically triggered when a UE changes its serving base station, and a contention-based random access, in which multiple UEs compete for channel access, potentially leading to collisions. The contention-based random access procedure in CIoT networks can follow a four-step or two-step random access procedure, 
    as illustrated in Figure~\ref{fig:steps_en}. The four-step contention-based random access procedure is detailed below. 

    In the first step, the UE randomly selects an RA preamble sequence from a set of available preambles for CBRA to be transmitted on the PRACH, referred to as Msg1. Collisions may occur at this stage if multiple UEs transmit the same preamble within the same random access opportunity (RAO); however, these collisions are not typically detectable by the base station at this stage.

    In the second step, the BS detects the preambles on the received PRACH signal and allocates the available uplink resources to those preambles.   This allocation is made through a Random Access Response (RAR), known as Msg2. The RAR is transmitted over the Physical Downlink Shared Channel (PDSCH) and contains a Random Access Network Temporary Identifier (RA-RNTI), which helps associate this response with the proper transmitted preamble and time slot. 

    In the third step, each UE that successfully receives the RAR proceeds to transmit the message triggering the RA procedure over the Physical Uplink Shared Channel (PUSCH), referred to as Msg3. This message includes the RA-RNTI and the UE's identity or a temporary identifier. 
    
    When multiple UEs select the same preamble within a given RA slot, they are assigned the same PUSCH resources and receive identical Random Access Responses (RARs), including the same RA-RNTI. As a result, they proceed to transmit Msg3 simultaneously on the same uplink resources. This potentially leads to an Msg3 collision at the BS.
    
    In the fourth step, the BS resolves contentions by sending a contention resolution message, known as Msg4, addressed to a single UE based on the Msg3 information. If a UE does not receive Msg4 within a specific time window, it assumes that the random access attempt has failed due to a collision and re-initiates the random access procedure by restarting from Msg1~\cite{UMTS, LTE1, LTE2, LTE3}. The complete four-step procedure requires two full round-trip signaling cycles between the UE and the BS. To improve efficiency and reduce access latency, the 3GPP standard introduces a two-step CBRA procedure~\cite{two}. 
    
    In the two-step approach, Msg1 and Msg3 are combined into a single message, termed MsgA, while Msg2 and Msg4 are merged into another message, termed MsgB. This modification effectively reduces the signaling to a single round-trip, which is particularly advantageous for ultra-reliable low-latency communications (URLLC) and NTN scenarios.

.



\section{Methodology}
\label{sec:metodo}

    This section outlines the methodology adopted to develop a machine learning-based framework for detecting collisions during the RA procedure in CIoT networks.
    
\begin{figure}[!t]
    \includegraphics[scale=0.20]{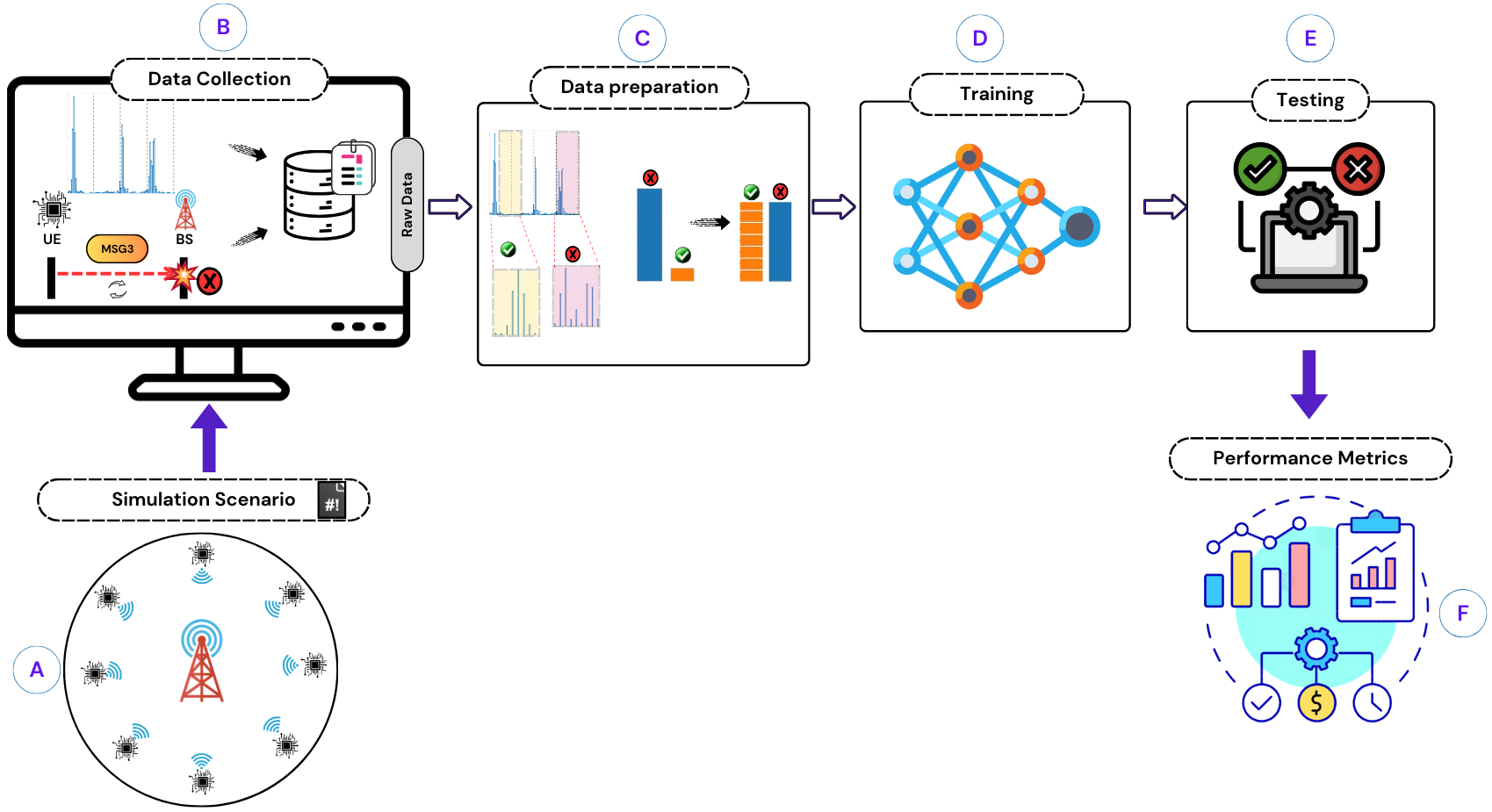}
    \centering
    \caption{Pipeline of the proposed machine learning framework for detecting preamble collisions in the RA procedure.}
    \centering
    \label{fig:model_process}
    \end{figure}

   The proposed approach utilizes  supervised classification of correlation patterns that are extracted from the PRACH signal at the BS after transmitting Msg1. Collision labels are assigned based on the outcomes observed following Msg3, which allows the model to be trained with ground-truth information that is not available during the reception of Msg1.  
   Figure~\ref{fig:model_process} illustrates the main stages of the pipeline: (i) dataset generation through realistic simulation of the RA process, (ii) preprocessing and class balancing, (iii) training and evaluation of various machine learning models, and (iv) model optimization via quantization techniques.


Each of these stages is described in detail in the following subsections.

\subsection{Dataset Generation}
\label{subsec:Dataset_Generation}

    A critical step for developing the proposed machine learning-based collision detection framework is the construction of a representative and realistic dataset. In this work, due to the lack of public datasets containing detailed PRACH signal information and preamble selection outcomes, we generated a proprietary dataset using MATLAB's LTE System Toolbox.

    The dataset generation process involves two main stages: the definition of simulation scenarios and parameters, and the simulation of random access events and preamble detection, capturing both the received signal characteristics and the occurrence of collisions.

    \subsubsection{Scenario definition and setup}

    We considered three distinct application scenarios, each designed to reflect realistic wireless communication environments with different mobility and channel conditions: (i) DS1: EPA (Extended Pedestrian A) model, 5 Hz Doppler frequency, 790 m cell radius (low mobility); (ii) DS2: EPA model, 5 Hz Doppler frequency, 500 m cell radius; (iii) DS3: ETU (Extended Typical Urban) model, 70 Hz Doppler frequency, 790 m cell radius (high mobility, rich multipath).   
    Additionally, the simulation includes 30,000 users (UE), 54 available preambles ($N_{\text{ZC}}$), two receiver antennas ($N_{\text{rx}}$), and 2,100 RAOs, among other network parameters such as the number of time slots and events. To model user access behavior, UE arrivals across time slots follow a beta distribution with parameters (3, 4), resulting in a temporally varying traffic load. The BS is positioned at the center of the cell, and UEs are randomly distributed throughout the coverage area. This spatial and temporal setup reflects the bursty and irregular access behavior observed in practical CIoT scenarios and serves as the foundation for the subsequent signal-level simulation.

 \subsubsection{RACH simulation and label assignment}

Based on the scenario configuration described in the previous subsection, we performed a time- and event-driven simulation of the RA procedure. For each time slot and event, the set of active UEs was determined, and their transmission behavior was simulated in detail.
Each UE was randomly positioned within the cell, and its distance to the BS was used to compute the corresponding propagation delay. This delay was converted into sample units based on the system’s sampling rate. To simulate realistic signal arrival times at the receiver, zero-padding was applied to each user’s signal, aligning it temporally with the computed delay.

The PRACH signals were then processed through a realistic wireless channel model, incorporating multipath fading and additive white Gaussian noise (AWGN). Signals from all active UEs were combined at the BS using Maximum Ratio Combining (MRC), and preamble detection was performed via correlation with a locally generated root ZC sequence. The resulting PDP was analyzed to identify peaks corresponding to received preambles. A dynamic detection threshold was set based on estimated noise power, and multiple correlation peaks within the same bin indicated the presence of potential collisions.

At this stage, although the BS can detect the preambles transmitted, it is still unaware of whether they result in successful access. To address this, we extended the simulation to include the third step of the RA procedure (Msg3 transmission). When two or more UEs selected the same preamble and were assigned the same uplink resources, their overlapping transmissions resulted in a collision at the BS. This behavior was used to assign binary labels to each PDP bin: non-collision (0) if the preamble was used by a single UE and successfully decoded; collision (1) if two or more UEs selected the same preamble in the same slot.

All simulation output was initially stored in a 1,200,640 × 27 matrix, representing all PDP bins across all RA events, including those without any detected transmissions. This matrix was then filtered to retain only the bins corresponding to actual transmissions, resulting in the final dataset with dimensions 579,533 × 28. Each row in the dataset contains the power values of the 24 PDP samples within a bin, metadata about the event and slot, and the assigned collision label.

\begin{figure}[!t]
    \includegraphics[scale=0.25]{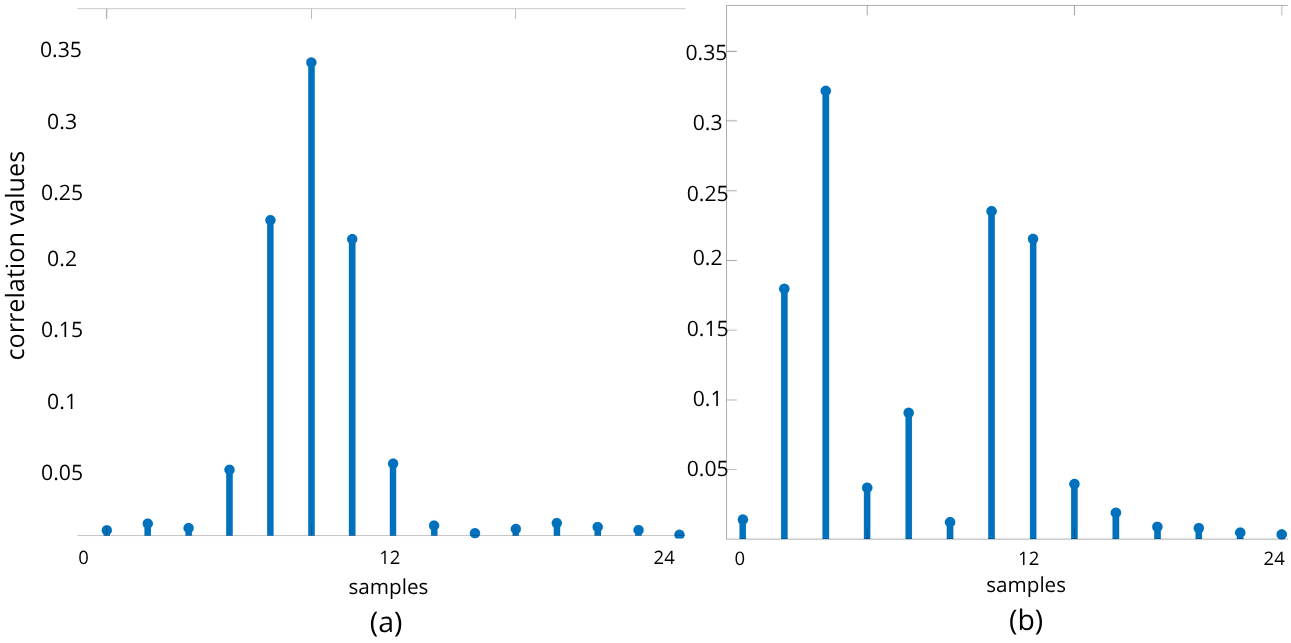}
    \centering
    \caption{ PDP bin examples for preamble detection: (a) non-collision (single peak); (b) collision (multiple peaks in the same ZCZ).
    }
    \label{fig:pdps}
    \end{figure}

Figure~\ref{fig:pdps} illustrates two examples of PDP bins: one with a single peak indicating a successful transmission, and another with two distinct peaks in the same ZCZ, revealing a preamble collision.

    \subsection{Preprocessing and Class Balancing}

   After the dataset was labeled based on the outcome of Msg3 transmissions, a substantial class imbalance was observed. The collision class (1) was significantly overrepresented with 472,369 samples (approximately 81.51\%), compared to the non-collision class (0), which had 107,164 samples (approximately 18.49\%). This imbalance can lead to biased models that tend to overpredict the majority class, reducing the model’s ability to correctly identify non-collision events and increasing the risk of false positives. In practical systems, such misclassification may result in inefficient use of radio resources, as unnecessary retransmissions can be triggered.

    \revtext{To mitigate this impact, we applied the SMOTE–Tomek Links technique~\cite{smotetomek}. SMOTE (Synthetic Minority Oversampling Technique)~\cite{smote} generates synthetic samples for the minority class by interpolating between existing samples and their nearest neighbors. Tomek Link~\cite{tomek}, in turn, removes borderline instances from the majority class that are close to the minority class, thereby reducing class overlap and noise. This combined approach balances the dataset more effectively than traditional resampling techniques such as Random Oversampling (ROS) \cite{ooversampling} or Random Sampling (RS) \cite{cochran}, and is particularly well-suited to our case, which consists exclusively of numerical features derived from signal data. As a result, the dataset was adjusted to contain 425,079 samples for each class, achieving a perfectly balanced 50\% distribution.}


    To mitigate this impact, we applied the SMOTE–Tomek Links technique~\cite{smotetomek}. SMOTE (Synthetic Minority Oversampling Technique)~\cite{smote} generates synthetic samples for the minority class by interpolating between existing samples and their nearest neighbors. Tomek Link~\cite{tomek}, in turn, removes borderline instances from the majority class that are close to the minority class, thereby reducing class overlap and noise. This combined approach balances the dataset more effectively than traditional resampling techniques such as Random Oversampling (ROS) \cite{ooversampling} or Random Sampling (RS) \cite{cochran}, and is particularly well-suited to our case, which consists exclusively of numerical features derived from signal data.

    \subsection{Collision Detection Mechanism Based on Machine Learning}

    Based on the labeled and balanced dataset described in the previous subsections, we implemented a supervised learning approach to detect preamble collisions during the RA procedure. The objective is to identify, from Msg1 reception alone, whether a given PDP bin corresponds to a collision event, thereby enabling more efficient PRACH resource allocation.

    The dataset consists of 28 columns, including: the event identifier, the RA slot number, the compartment index, 24 power values from a PDP compartment, and a binary label. This label indicates the presence (1) or absence (0) of a collision based on the outcome of Msg3. The dataset contains 579,533 observations. We reserved 90\% of the data for model development and 10\% for final testing. During training, a portion of the training set was used for validation to tune model hyperparameters.

    We evaluated the performance of several classification models, including: Logistic Regression, Support Vector Machine (SVM), Decision Tree (DT), Random Forest, Naive Bayes, K-Nearest Neighbors (KNN), LightGBM, XGBoost, and a Neural Network (NN). These models span different learning paradigms and exhibit varying levels of complexity, interpretability, and computational demand. 

Linear models, such as logistic regression, are simple, fast, and interpretable. However, they struggle with complex non-linear patterns. Probability-based models, like Naive Bayes, are efficient and well-suited for high-dimensional problems, although their independence assumption can limit accuracy. 

Tree-based methods, including decision trees, random forests, XGBoost, and lightGBM, effectively capture non-linear relationships and interactions. The ensemble versions of these models achieve high predictive power but sacrifice interpretability and increase computational demands. 

Distance-based approaches, such as K-Nearest Neighbors (KNN), are intuitive and flexible. However, they become inefficient with large datasets. 

Margin-based models, like Support Vector Machines (SVMs), excel in high-dimensional spaces and can accommodate non-linearities through the use of kernels. Nevertheless, they are computationally intensive and less interpretable. 

Finally, neural networks are distinguished by their ability to model highly complex, non-linear relationships across diverse data types. However, require large datasets, significant computational power, and careful tuning. Additionally, they often act as ``black boxes", making interpretation challenging.

This diverse selection allows us to explore trade-offs between detection performance, interpretability, and computational efficiency, aiming to identify the most suitable model for real-time collision detection in dense CIoT deployments. Note that only classic and simple NN models were considered due to the stringent inference-time requirements of mobile communication technologies.

    \subsection{Model Optimisation by Quantization Techniques}

    To enable real-time inference on resource-constrained devices, we applied quantization techniques to compress the trained models and improve inference efficiency. Initially, the models were saved in \texttt{.h5} format using 32-bit floating-point (float32) precision, which preserves the full range of weights and activations but results in large file sizes and slow inference performance.

    To address these limitations, we explored two
    quantization strategies: Dynamic Range Quantization (DRQ) and Full Integer Quantization (FIQ). DRQ converts only the model weights to 8-bit integers (int8), while keeping activations in float32. This method is simple to implement and does not require a calibration dataset, offering a good trade-off between model size and accuracy \cite{dynamic}.
    In contrast, FIQ converts both weights and activations to int8. This technique requires a representative calibration dataset to accurately map the floating-point ranges into integer values. Although more complex, it achieves greater reductions in inference latency and memory usage, making it particularly suitable for scenarios that demand high performance on resource-constrained devices \cite{full_integer}.
    
    Quantization can lead to a loss of precision due to the reduced numeric representation of weights and activations. To mitigate this effect, we applied post-training quantization using a representative calibration dataset, which helped preserve the model’s predictive accuracy during inference. This ensures that the quantized model remains robust and generalizable across a variety of CIoT scenarios, including those with high user mobility.

    \begin{figure}[!t]
    \includegraphics[scale=0.25]{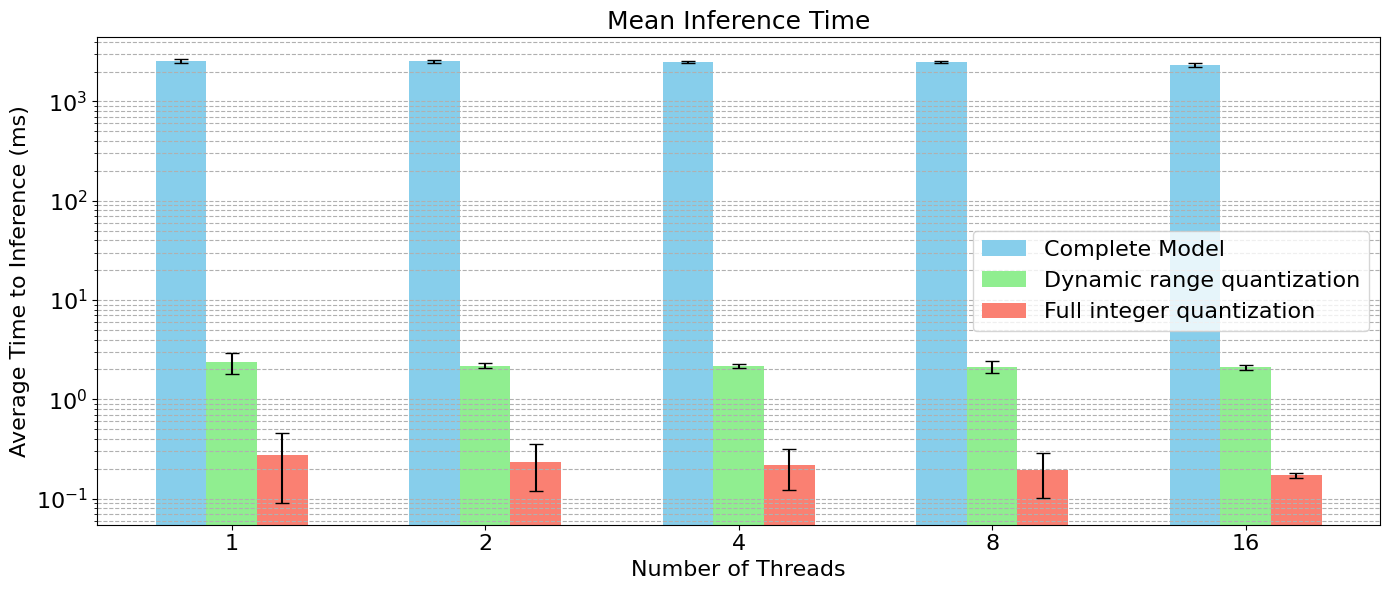}
    \centering
    \caption{Mean inference time for quantized and non-quantized models across different thread configurations.}
    \centering
    \label{fig:mean_time}
    \end{figure}
\def\mOne{S1}   
\def\mTwo{S2}   
\def\mThree{S3} 
\def\mFour{S4}  
\begin{table*}[!t]
    \centering
    \caption{Classification metrics for all models and considered scenarios.}
    \label{tab:performanc_metrics}
    \begin{adjustbox}{width=1\textwidth}
        \begin{tabular}{lcccccccccccccccc}
            \hline
            
            \multirow{3}{*}{\centering\textbf{Model}} & \multicolumn{16}{c}{\textbf{Performance Metrics}} \\
            & \multicolumn{4}{c}{Precision} & \multicolumn{4}{c}{Recall} & \multicolumn{4}{c} {Specificity} & \multicolumn{4}{c}{Balanced Accuracy}
 \\
            & \mOne & \mTwo & \mThree & \mFour &
            \mOne & \mTwo & \mThree & \mFour & \mOne & \mTwo & \mThree & \mFour & \mOne & \mTwo & \mThree & \mFour \\
            \midrule
            Logistic Regression & 0.9714 & 0.9621 & 0.9718 & 0.8682
            & 0.9985 & 0.9981 & 0.9980 & 0.9999 
            & 0.8875 & 0.8563 & 0.8888 & 0.6335  
            & 0.9430 & 0.9272 & 0.9434 & 0.8167 \\ 
            Support Vector Machines & 0.9716 & 0.9626 & 0.9719 & 0.8745
            & 0.9988 & 0.9983 & 0.9983 & 0.9999
            & 0.8883 & 0.8582 & 0.8893 & 0.6447 
            & 0.9435 & 0.9283 & 0.9438 & 0.8223 \\
            Decision Trees & 0.9735 & 0.9737 & 0.9740 & 0.9447  
            & 0.9802 & 0.9805 & 0.9794 & 0.9842
            & 0.8868 & 0.8882 & 0.8882 & 0.7936 
            & 0.9335 & 0.9343 & 0.9338 & 0.8889 \\
            
            Random Forest & 0.9756 & 0.9754 & 0.9757 & 0.9545  
            & 0.9979 & 0.9979 & 0.9971 & 0.9998
            & 0.9023 & 0.9020 & 0.9022 & 0.8334 
            & 0.9501 & 0.9499 & 0.9496 & 0.9166 \\
            Naive Bayes & 0.8180 & 0.8166 & 0.8703 & 0.6454  
            & 0.9923 & 0.9923 & 0.9942 & 0.9974
            & 0.5480 & 0.5471 & 0.6312 & 0.3894  
            & 0.7702 & 0.7697 & 0.8127 & 0.6934 \\
            K-Nearest Neighbor & 0.9756 & 0.9753 & 0.9755 & 0.9410 
            & 0.9999 & 0.9998 & 0.9998 & 0.9998
            & 0.9031 & 0.9023 & 0.9027 & 0.7944 
            & 0.9515 & 0.9510 & 0.9512 & 0.8971 \\
            LightGBM & 0.9754 & 0.9743 & 0.9754 & 0.9498 
            & 0.9995 & 0.9990 & 0.9991 & 0.9999
            & 0.9021 & 0.8983 & 0.9022 & 0.8194 
            & 0.9508 & 0.9486 & 0.9507 & 0.9097 \\
            XGBoost & 0.9756 & 0.9748 & 0.9755 & 0.9553 
            & 0.9992 & 0.9988 & 0.9990 & 0.9997
            & 0.9029 & 0.9000 & 0.9024 & 0.8361  
            & 0.9511 & 0.9494 & 0.9507 & 0.9179 \\
            Neural Network & 0.9999 & 0.9994 & 0.9997 & 0.9999  
            & 0.9755 & 0.9726 & 0.9749 & 0.9070 
            & 0.9996 & 0.9977 & 0.9987 & 0.9999 
            & \textbf{0.9876} & \textbf{0.9851} & \textbf{0.9868} & \textbf{0.9535} \\
            
            \bottomrule
        \end{tabular}
      \label{tab:table1}    
    \end{adjustbox}
\end{table*}
   
    The inference process consists of generating predictions on new data using either the original model (saved in \texttt{.h5} format) or its quantized version (in \texttt{.tflite} format), depending on the deployment context. The \texttt{.tflite} format is particularly advantageous for embedded devices due to its reduced size and faster loading time.
    During evaluation, input samples composed of 24 features were processed using the TensorFlow or TensorFlow Lite interpreter. To ensure stable and unbiased measurements, we performed warm-up inferences before benchmarking. Inference time was measured using a high-precision timer across 64 dataset samples. For each run, we recorded the total duration and computed both the average inference time and its standard deviation.

    This quantization and benchmarking procedure establishes the computational feasibility of the proposed solution, ensuring it meets the constraints of real-world CIoT environments while maintaining reliable performance.

\section{Performance Evaluation}
\label{sec:Performance}

\subsection{Experimental setup}
\label{sec:setup}

The performance of the proposed mechanism for preamble collision detection was assessed using nine classical machine learning models: Logistic Regression, SVM, Decision Tree, Random Forest, Naive Bayes, KNN, LightGBM, XGBoost, and NN. Each model was trained and evaluated on datasets generated under four experimental scenarios, as described in Section~\ref{sec:metodo}, encompassing variations in channel conditions, coverage, and mobility profiles.

The four experimental scenarios are summarized as follows:
\begin{itemize}
\item \textbf{S1:} Training and testing on the EPA channel model (5 Hz Doppler shift, 790 m user radius).
\item \textbf{S2:} Training and testing on the ETU channel model (70 Hz Doppler shift, 790 m user radius).
\item \textbf{S3:} Training on EPA (790 m), testing on EPA with reduced coverage (500 m).
\item \textbf{S4:} Training on EPA (790 m), testing on ETU (70 Hz, 790 m).
\end{itemize}

Scenarios S1 and S2 correspond to intra-scenario evaluations, where training and testing are performed under identical channel conditions. In contrast, S3 and S4 introduce cross-scenario generalization challenges, involving either coverage changes or mismatched channel models between training and testing phases.

These cross-scenario evaluations are particularly important for assessing the robustness of the models beyond the training domain. In real-world CIoT deployments, devices may operate under varying propagation conditions, such as reduced coverage areas, increased user mobility, or different multipath profiles, not seen during training. By simulating these mismatches, S3 and S4 allow us to evaluate how well each model generalizes to unseen conditions, identify potential overfitting to specific channel configurations, and ultimately select models with greater adaptability for deployment in dynamic and heterogeneous environments.

\subsection{Performance Evaluation Metrics}
For evaluating model performance, we focus on the balanced accuracy as our target metric. 
Unlike the traditional accuracy metric, the Balanced Accuracy includes the Specificity, which is critical for our particular application. Specificity captures the ability of the model to correctly recognize non-collision events, thereby reducing false alarms. The rationale for this is that a non-collided preamble will adequately utilize the resources of the network in the subsequent phases of the RA procedure. For instance, if the proposed mechanism is used to avoid allocation of resources to those preambles detected as collisions, we are interested in the model that predicts with high precision this type of event.  
Moreover, this metric provides a more reliable assessment by accounting for the model's performance across both classes. 
Thus,
for completeness, we also report Precision, Recall, and Specificity. 

Moreover, average inference latency is also considered to assess the practical applicability of the best-performing model under realistic hardware and timely constraints of typical CIoT protocols and implementations.

\subsection{Intra-Scenario Evaluation (S1 and S2)}
\label{sec:Results_discussions}


In Scenarios S1 and S2, where propagation conditions remain consistent between training and testing, most models achieved Balanced Accuracy scores above 90\%. As shown in Table~\ref{tab:table1}, the Neural Network consistently outperformed the other models, reaching 98.76\% in S1 and 98.51\% in S2. These results highlight its ability to capture complex, non-linear relationships in the input features, an essential trait for accurate collision detection across diverse mobility profiles.

\revtext{Beyond Balanced Accuracy, nearly all models exhibited high Recall (typically above 0.98), indicating strong sensitivity in detecting collision events. However, Specificity varied more significantly. For instance, Naive Bayes achieved Recall close to 0.99 but much lower Specificity (around 0.55), suggesting a marked tendency to overpredict collisions. This results in a high rate of false positives, which undermines its overall effectiveness despite its apparent detection strength. By contrast, the Neural Network combined high Recall with near-perfect Specificity (0.9996 in S1 and 0.9977 in S2), yielding highly balanced performance. This symmetry is especially valuable in practical massive IoT settings, where excessive false positives can lead to unnecessary retransmissions and inefficient use of resources. Other ensemble models, such as Random Forest, XGBoost, and LightGBM, also achieved strong results but did not match the Neural Network’s consistent performance across both classes.}

Naive Bayes, on the other hand, was the worst-performing model in both scenarios, with Balanced Accuracy of 77.02\% in S1 and 76.97\% in S2. This poor performance is likely due to its strong assumption of feature independence, which does not hold in the context of correlated PDP power components. Furthermore, a slight performance degradation observed across all models from S1 to S2 can be attributed to the more complex nature of the ETU channel, which introduces higher Doppler spread and richer multipath, increasing variability in the input features.

\subsection{Cross-Scenario Generalization (S3 and S4):}

Scenarios S3 and S4 evaluate the generalization capability of the models when test data diverge from training conditions. In Scenario S3, models are trained on the EPA channel with a 790 m radius but tested on the same channel type with reduced coverage (500 m), simulating spatial variability. Scenario S4 presents a more challenging setting: models trained on the EPA channel are tested on ETU, introducing a mismatch in propagation characteristics and mobility profiles. Despite this shift, most models performed consistently well in S3. Balanced Accuracy declined only marginally compared to S1, indicating that reduced coverage had limited impact on feature distributions. Decision Trees, Random Forest, and LightGBM maintained scores above 93\%, while the Neural Network reached 98.68\%, confirming strong robustness to spatial variation.

Scenario S4, by contrast, resulted in more pronounced performance degradation. The higher Doppler spread and richer multipath of the ETU channel increased signal variability, challenging generalization. Several models exhibited substantial drops, for instance, Logistic Regression and Naive Bayes fell to 81.67\% and 69.34\%, respectively. Nevertheless, four models proved resilient: Random Forest (91.66\%), LightGBM (90.97\%), XGBoost (91.79\%), and the Neural Network, which maintained a standout performance of 95.35\%. These results confirm the Neural Network’s superior ability to generalize under mismatched channel and mobility conditions. While ensemble models such as XGBoost and LightGBM also demonstrated strong generalization, the Neural Network remained the most consistent across all scenarios. Its resilience under both matched and mismatched conditions further justifies its choice as the basis for deployment optimization.

\subsection{Inference-time Evaluation}

Based on its consistently superior performance across all scenarios and a well-balanced trade-off between Recall and Specificity, the Neural Network was chosen as the target model for deployment optimization. 

To enhance its suitability for real-time, resource-constrained environments, post-training quantization was applied using two strategies: DRQ and FIQ.

Figure~\ref{fig:mean_time} compares the inference latency of the original and quantized models. While the original model exhibited inference times around 2500 ms, DRQ reduced this to approximately 2 ms, and FIQ further decreased it to just 0.3 ms, representing improvements of several orders of magnitude.

Table~\ref{tab:comparison} presents a detailed comparison of classification metrics for the original and quantized Neural Network models. All variants-original, DRQ, and FIQ-achieved nearly identical Precision, Recall, Specificity, and Balanced Accuracy, with differences below 0.0003. Notably, Full Integer Quantization preserved a Balanced Accuracy of 98.67\%, closely matching the original model’s 98.70\%. This negligible degradation confirms that, when properly calibrated, quantization maintains model quality.
The computational efficiency gains, however, are substantial. As illustrated in Figure~\ref{fig:mean_time}, FIQ achieves ultra-low inference latency across all thread configurations, outperforming both the original and DRQ models. The use of \texttt{int8} weights and activations enables faster execution and a reduced memory footprint, making FIQ highly suitable for deployment in latency-sensitive edge applications.

\begin{table}[!t]
    \centering
    \caption{Classification performance of non-quantized and quantized neural network models.}
    \label{tab:comparison}
    \small 
    \setlength{\tabcolsep}{5pt} 
    \renewcommand{\arraystretch}{1.1} 
    \begin{tabular}{lccc}
    \toprule
    \multirow{2}{*}{\textbf{Performance Metric} } & 
    \multicolumn{3}{c}{\textbf{Quantization Technique}} \\ 
    & None & Dynamic Range & Full Integer \\ 
                               \midrule
    Precision           & 0.9997                 & 0.9997                 & 0.9997                \\ 
    Recall              & 0.9749                 & 0.9749                 & 0.9741                \\ 
    Specificity        & 0.9990                 & 0.9990                 & 0.9990                \\ 
    Balanced Accuracy  & 0.9870                 & 0.9870                 & 0.9867                \\ \bottomrule
    \end{tabular}

\end{table}





    
    


\section{Conclusion}
\label{sec:Conclusion}
   This paper proposed a machine learning-based mechanism for early collision detection in the Random Access procedure of Cellular IoT networks. By generating a labeled dataset under realistic LTE and 5G scenarios, we systematically evaluated nine classification models across diverse propagation, mobility, and coverage conditions. The neural network consistently outperformed the other models, achieving over 98\% Balanced Accuracy in in-distribution evaluations and maintaining 95\% in out-of-distribution scenarios, thus demonstrating robustness to varying network scenarios.

   To ensure feasibility in real-world deployments, we applied post-training quantization techniques. Full Integer Quantization reduced inference time from 2500 ms to 0.3 ms with negligible accuracy loss, enabling efficient execution on base station hardware. These results confirm that the proposed approach combines high detection accuracy with ultra-low latency inference, making it a practical solution for large-scale, real-time CIoT and mMTC applications.

\bibliographystyle{IEEEtran}
\bibliography{References}

\end{document}